\documentclass[prl,aps,twocolumn,longbibliography]{revtex4-1}
\usepackage{color}
\usepackage{graphicx}
\usepackage{bm}
\usepackage{amsmath}
\usepackage{amssymb}
\usepackage{amsthm}
\usepackage{amsfonts}
\usepackage{enumerate}
\usepackage{latexsym}

\DeclareMathOperator{\var}{var}
\DeclareMathOperator{\cov}{cov}

\usepackage{braket}

\usepackage{dcolumn}
\usepackage{amsfonts}
\usepackage{enumerate}
\newcolumntype{P}[1]{>{\centering\arraybackslash}p{#1}}
\usepackage{float}
\newcommand\widebar[1]{\mathop{\overline{#1}}}

\DeclareMathOperator{\IM}{Im}

\usepackage{array}

\usepackage[colorlinks=true,urlcolor=blue,citecolor=blue,allcolors=blue]{hyperref}

\usepackage{float}
\usepackage{cleveref}

\begin{document}

\title{Criticality in amorphous topological matter --- beyond the universal scaling paradigm}
\author{Moein N.~Ivaki, Isac Sahlberg, and  Teemu Ojanen}
\affiliation{Computational Physics Laboratory, Physics Unit, Faculty of Engineering and
Natural Sciences, Tampere University, P.O. Box 692, FI-33014 Tampere, Finland}
\affiliation{Helsinki Institute of Physics P.O. Box 64, FI-00014, Finland}

\begin{abstract}
We establish the theory of critical transport in amorphous Chern insulators and show that it lies beyond the current paradigm of topological criticality epitomized by the quantum Hall transitions. We consider models of Chern insulators on percolation-type random lattices where the average density determines the statistical properties of geometry. While these systems display a two-parameter scaling behaviour near the critical density, the critical exponents and the critical conductance distributions are strikingly nonuniversal. Our analysis indicates that the amorphous topological criticality results from an interpolation of a geometric-type transition at low density and an Anderson localization-type transition at high density. Our work demonstrates how the recently discovered amorphous topological systems display unique phenomena distinct from their conventionally-studied counterparts.

\end{abstract}
\maketitle

\emph{Introduction.---}Recent theoretical advances have brought the full topological classification of crystalline matter in sight~\cite{vergniory2019complete,zhang2019catalogue,tang2019comprehensive}. However, there are rapidly emerging lines of research in topological systems without spatial symmetry. Since nontrivial topology in general does not rely on spatial order, amorphous systems provide an interesting new platform for topological matter~\cite{li2020large,zhou2020amorphous,corbae2019evidence,mitchell2018amorphous,mahendra2018room,costa2019toward,yang2019metal,mukati2020topological,marsal2020topological,agarwala2020higher,mitchell2018amorphous,poyhonen2018amorphous,Agarwala2017prl,fulga2014statistical,PhysRevB.96.100202}. Previously, the question as to whether the topological behaviour of amorphous systems and crystalline systems display fundamental differences has remained largely unclear. In this work we answer this question affirmatively by establishing that the critical transport of amorphous Chern insulators exhibit striking departures from their spatially-ordered counterparts.  

The theory of quantum Hall (QH) plateau transitions, initiated by Khmelnitskii and Pruisken~\cite{khmel1983quantization,pruisken1985dilute}, has achieved a paradigmatic role in the theory of topological phase transitions. This theory, with generalizations to various symmetry classes and models, describes topological phase transitions as a form of Anderson localization (AL) transition with diverging localization length (LL)~\cite{huckestein1995scaling,evers2008anderson}. The topological phase transition corresponds to an unstable fixed point, characterized by universal critical exponents, in a two-parameter space. This picture, with appropriate modifications, is believed to capture the generic features of topological phase transition in non-interacting systems. In particular, the transitions are classified by a set of universal critical exponents that only depend on the symmetries and generic features of the system but not on microscopic details. While theoretically predicted values for the LL exponents in the QH transition exhibit a degree of variation and seem to somewhat overestimate the experimental ones~\cite{koch1991size,engel1993microwave,wei1988experiments,li2005scaling,chang2016observation,li2009scaling}, the values extracted from widely different models typically fall between $\nu=2.4-2.6$~\cite{amado2011numerical,puschmann2019integer,zhu2019localization,klumper2019,gruzberg2017geometrically,obuse2012finite,slevin2009critical,obuse2010conformal,zhu2019localization,fulga2011topological,huo1992current}. This degree of agreement lends significant credibility to the orthodox theory. 

\begin{figure}[t]
\includegraphics[width=0.95\columnwidth]{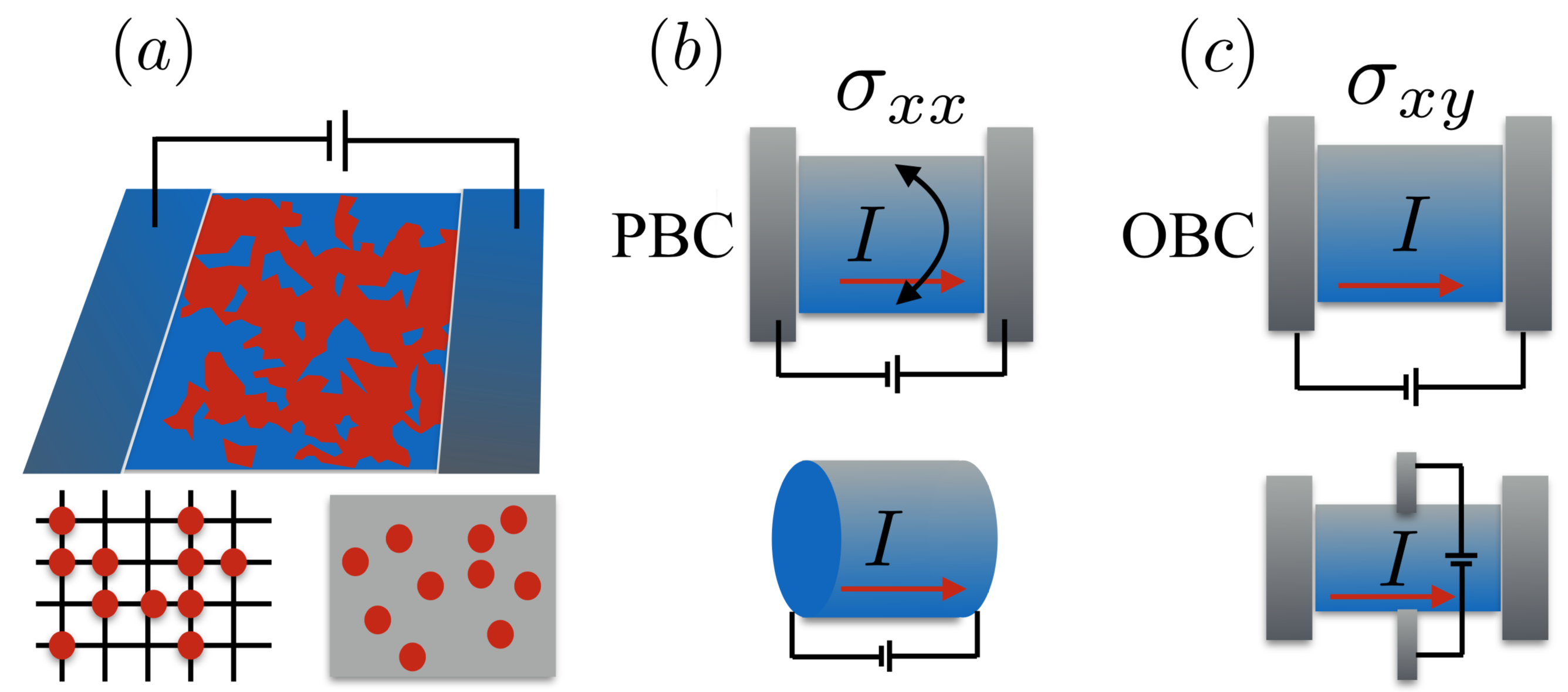}
\caption{(a) Schematic setup for transport studies in amorphous Chern insulators. The studied random geometries are generated by discrete and continuum percolation lattices. (b) Longitudinal conductivity can be extracted from the two-terminal conductance with periodic boundary conditions in the transverse direction (top) which is equivalent to the setup below. (c) Hall conductivity can be extracted from the two-terminal setup with open boundary conditions (top). The conductivity corresponds to the Hall conductivity obtained from the four-terminal setup (bottom).}
\label{fig:cond_setup}
\end{figure}

\begin{figure*}[ht]
    \centering
    \includegraphics[width=1.0\linewidth]{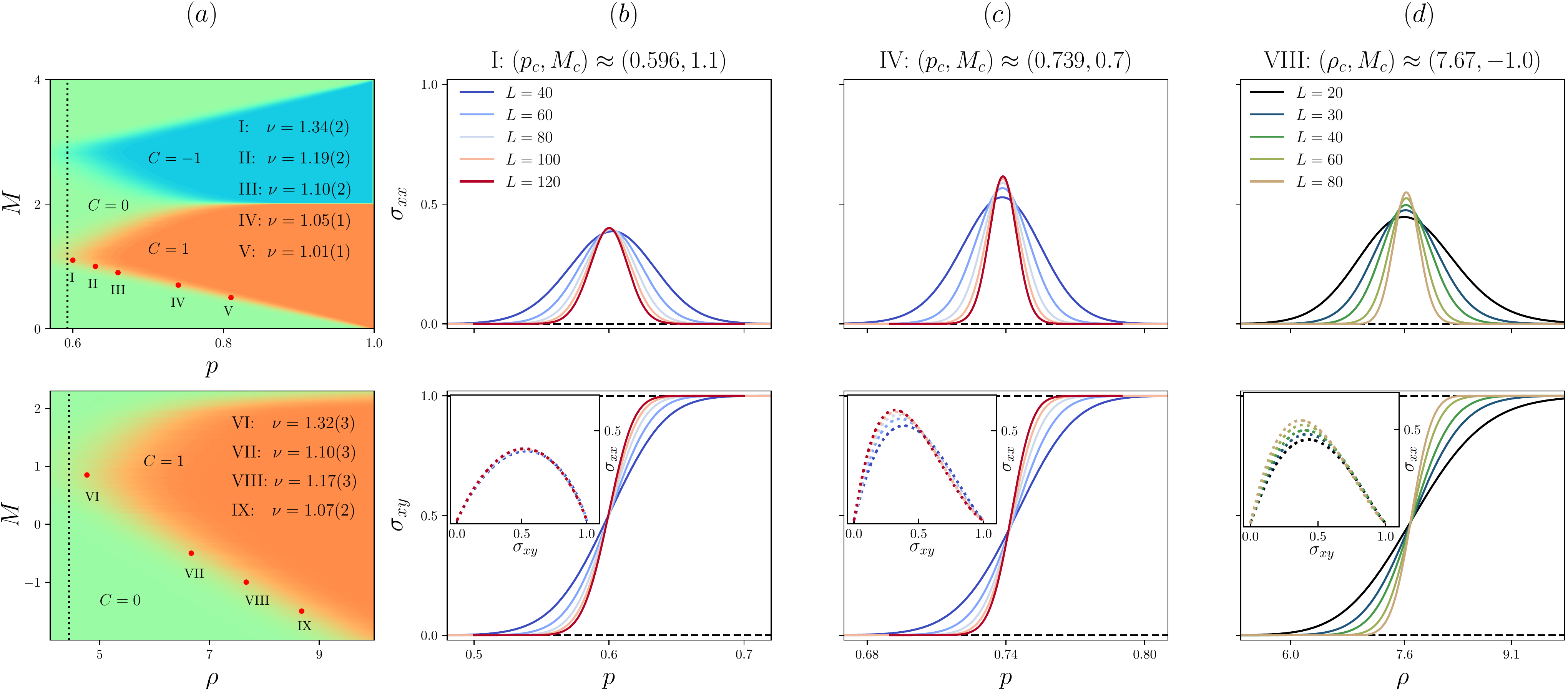}
    \caption{ (a): Topological phase diagram (Chern numbers) in the density-mass plane for the discrete (top) and the continuum (bottom) model. The red dots labelled by roman numerals indicate the positions where the scaling analysis was carried out. The black dotted line indicates the percolation threshold of the lattice. (b): Conductance scaling in the lattice model at the optimal point I. The inset in the bottom shows the flow in the conductivity plane. (c): Same as (b) but in a higher-density regime. (c): Conductance scaling in the continuum model at VIII. The curves are generated from over $10
   ^5$ configurations. }
    \label{fig:phdia_cond_sca}
\end{figure*}
In the present work we establish the critical theory of amorphous Chern insulators and show that it lies strikingly beyond the universal scaling paradigm. We study transport properties of amorphous topological states defined on random lattices with variable density as depicted in Fig.~\ref{fig:cond_setup}(a). By numerically evaluating configuration-averaged longitudinal and Hall conductivities $\sigma_{xx}$, $\sigma_{xy}$ in setups illustrated in Figs.~\ref{fig:cond_setup}(b) and (c), we study their scaling behaviour as a function of density. While conductivities are shown to obey two-parameter scaling behaviour near the critical density $\rho_c$, the critical exponent $\nu$ characterizing the diverging LL as $\xi \propto|\rho-\rho_c|^{-\nu}$ is strongly nonuniversal $\nu=1.01(1)-1.35(2)$. To further characterize the nonuniversality, we calculate the critical conductance distributions (CDs) and show how they interpolate between two distinct types, one which exhibits QH-type features at high density, and another which exhibits a striking low-conductance peak stemming from geometric fluctuations at low density. We conclude that the amorphous topological criticality (ATC) arises from the interpolation of a geometric percolation-type and the AL-type transitions.

\emph{Models of amorphous Chern insulators.---}Following Ref.~\cite{sahlberg2020topological}, we study two-band Chern insulators with the tight-binding Hamiltonian
\begin{equation}
H=\begin{pmatrix}\label{h1}
(2-M)\delta_{ij}+T_{ij} & iT_{ij}e^{-i\phi_{ij}}  \\
iT_{ij}e^{i\phi_{ij}} 	& -(2-M)\delta_{ij}-T_{ij} 
\end{pmatrix},
\end{equation}
where $M$ is the time-reversal breaking mass term in the units of a characteristic hopping amplitude and $T_{ij}=-\frac12 e^{-r_{ij}/\eta}\theta(R-r_{ij})$ describes the spatial decay of the hopping amplitudes. Here $r_{ij}=|\mathbf{r}_i-\mathbf{r}_j|$ is the distance between sites $i,j$, the parameters $\eta, R$ describe the decay of hopping, and the phase factor is given by $e^{i\phi_{ij}}=\frac{r_{ij}^x+ir_{ij}^y}{r_{ij}}$, where $r_{ij}^x = x_i-x_j$.  We mainly consider disc hopping models with $\eta=\infty$ but also check that the discovered qualitative features are present for smooth spatial decay with constant $\eta$ and $R\to \infty$. 

We study the model \eqref{h1} on random percolation-type geometries on a square lattice as well as in continuum as illustrated in Fig.~\ref{fig:cond_setup}(a). As in percolation theory, the lattice sites in the discrete case are independently populated with probability $p$, whereas in the continuum problems the sites are independently distributed in the 2d continuum with intensity $\rho$ particles per unit area. 

\begin{figure*}[ht]
    \centering
    \includegraphics[width=1.\linewidth]{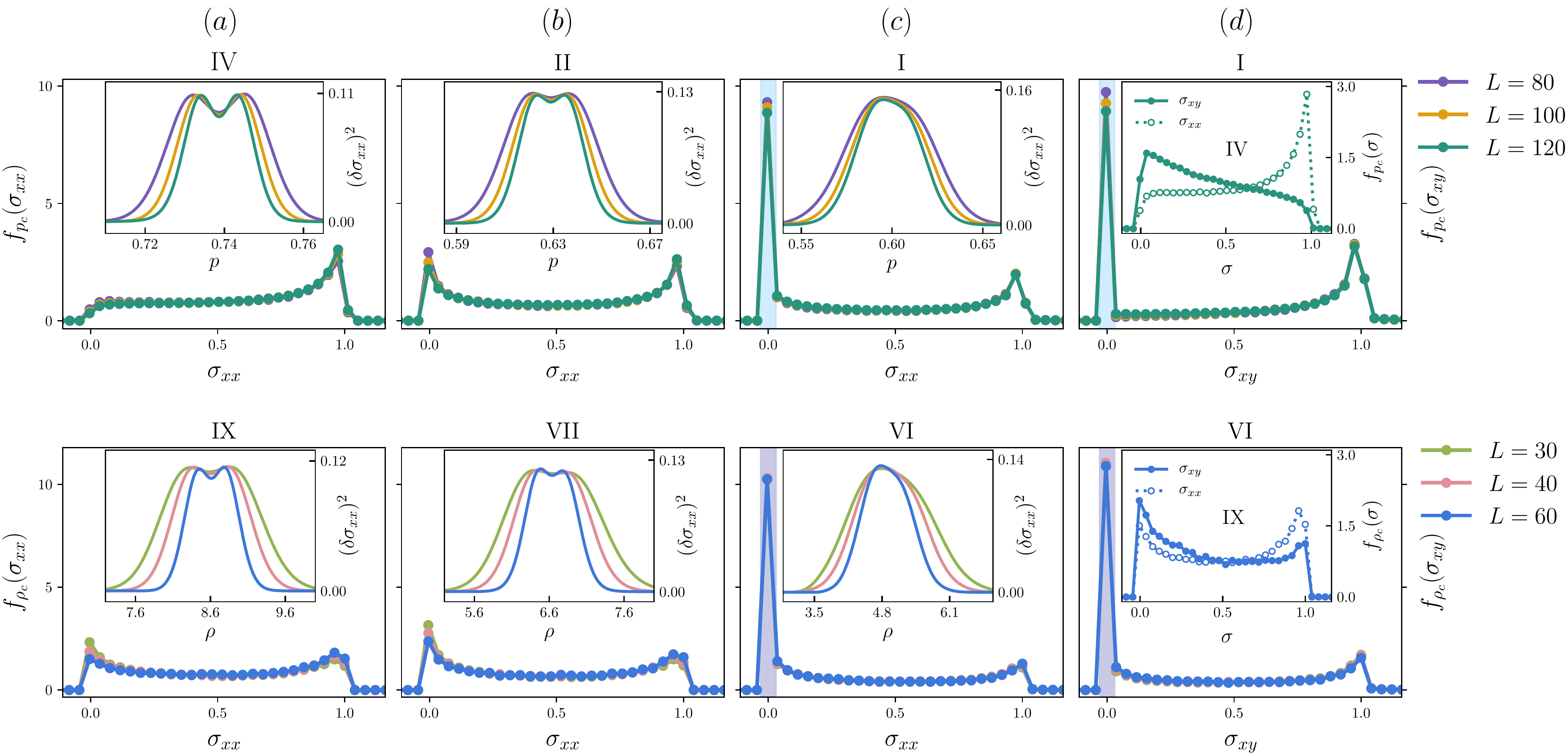}
    \caption{Evolution of CDs for the discrete (top row) and the continuum model (bottom row) along the phase boundary. The inset in (a)-(c) illustrates the variance of the distribution near critical density. The inset in (d) highlights the difference of the $\sigma_{xx}$ and $\sigma_{xy}$ distributions at high density. Distributions are generated from up to $10^5$ configurations.}
    \label{fig:dists}
\end{figure*}

\emph{Scaling theory of transport.---}We assume that the electronic states are half-filled (one electron per site) and study electrical conductance averaged over different random configurations as a function of the density of lattice sites. More precisely, in the discrete case we study the topological criticality as a function of $p$ and in the continuum case as a function of $\rho$. For discrete random realizations, we evaluate conductances by employing the KWANT package~\cite{groth2014kwant}. For continuum configurations, we employ the Green's function method outlined in Sec.~I of the Supplemental Information (SI). The longitudinal and Hall conductivities are obtained from square-shaped samples in the two-terminal setups illustrated in Figs.~\ref{fig:phdia_cond_sca}(b) and (c). The correspondence between the Hall conductivity obtained from the four-terminal setup and the two-terminal setup is illustrated in Sec.~III in the SI. The central piece of computational technology in our work is to carry out the configuration averages with fixed number of lattice sites $n$ and subsequently exploit the analytical connection between $n$ and $p$ ($\rho$). In Sec.~I of the SI we show that this procedure significantly reduces the statistical fluctuations compared to direct sampling of $p$.   

We postulate that the conductivities for density-driven topological transition satisfy a two-parameter scaling form
\begin{equation}\label{scale}
\sigma= F[L^{1/\nu}\zeta_1(p),L^{y}\zeta_2(p)],
\end{equation}
where $L$ is the linear system size, $\nu$ is the critical exponent of localization length, and $y<0$ describes the irrelevant scaling direction. Here $F(x,y)$ is an a priori unknown scaling function and $\zeta_1(p),\zeta_2(p)$ describe the relevant and irrelevant scaling variables, respectively. In the large system limit we recover a single-parameter scaling characterized by the LL $\xi\propto|p-p_c|^{-\nu}$ where $p_c$ is the critical density. For continuum problems we postulate a similar expression with $p$ and $p_c$ substituted by particle intensity per unit area $\rho$ and its critical value $\rho_c$. The statistical analysis of extracting $\nu$, $p_c$ and $y$ from the conductance data is presented in Sec.~II in the SI.  

The topological phase diagrams of lattice and continuum disc hopping models are evaluated following Ref.~\cite{sahlberg2020topological} and illustrated in Fig.~\ref{fig:phdia_cond_sca}(a). The red dots indicate the points I-V and VI-IX on the phase boundary where the critical parameters have been evaluated. The localization exponents are listed in Fig.~\ref{fig:phdia_cond_sca}(a) and the full scaling data is presented in Table I to III in the SI. The behaviour of conductivities as a function of density is illustrated in Figs.~\ref{fig:phdia_cond_sca}(b)-(d). In general, we obtain an excellent fit of the conductance data with the two-parameter scaling form at each studied point. For discrete and continuum disc models we observe that the nontrivial phase reaches down to the percolation threshold which is the theoretical lower limit for the topological phase for these models~\cite{sahlberg2020topological}. The critical density $p_c=0.596(2)$ at point I matches well the percolation threshold of square lattice $p^{cl}_c\approx0.593$. Also, the LL exponents at peak points I ($\nu=1.34(2)$) and VI ($\nu=1.32(3)$) are in excellent agreement with the correlation length exponent $4/3$ of 2d percolation \cite{stauffer2018introduction}. These results together indicate that when the critical density approaches the geometric percolation threshold of the lattice, the critical wave functions are restricted only by the geometry of the underlying lattice, not quantum interference effects.

At higher densities away from I and VI, the critical exponents do not agree with the low-density value and show large nonuniversal variation. This remarkable behaviour is in \emph{in striking contrast} to the universal behaviour of the disordered systems. In the studied regime we observe continuous variation of critical exponents $\nu=1.01(1)-1.35(2)$ for the discrete model and similar for the continuum disc model. As listed in the SI, the critical conductance values also exhibit large non-universal variation. This is in sharp contrast to QH systems, where the universality of $\sigma^c_{xx}$~\cite{wang1998scaling,wang1996critical,xue2013quantum,werner2015anderson} is believed to follow from universal multifractal properties~\cite{schweitzer2005universal,obuse2010conformal,evers2008multifractality,obuse2008boundary}. The strong variation of the critical properties suggests that topological phase transitions at high and low density regime are dominated by qualitatively different mechanisms. At low density, the agreement of $\nu$ and $p_c$ with the correlation length exponent and the threshold in classical percolation suggest that the reduced lattice connectivity drives the transition. The conductance distribution functions calculated below confirm this observation as well as suggest that the transition at high densities is dominated by conventional AL mechanism.

We note that the topological transition can also be induced at fixed density by varying the mass parameter $M$ through a critical point $(p_c,M_c)$ on a phase boundary. Since conductance is an analytic function of $p$ and $M$ for finite systems, the exponent $\nu'$ characterizing the divergence $\xi_M\propto|M-M_c|^{-\nu'}$ is expected to coincide with the critical exponent $\nu$ in the density-driven transition. Indeed, in Sec.~V in the SI we illustrate that the two exponents are consistent.  

\emph{Critical conductance distributions.---}To gain better insights into the critical behaviour, we now study the critical conductance distribution functions (See Sec.~III in the SI for technical details). Figs.~\ref{fig:dists}(a)-(c) illustrate the behaviour of the longitudinal CDs at I, II ,IV (top row) and VI, VII, IX (bottom row) indicated in Fig.~\ref{fig:phdia_cond_sca}. At high densities, distributions are qualitatively similar to the one shown in Fig.~\ref{fig:dists}(a), illustrating that the conductance is broadly distributed between 0 and 1 (in the units of $e^2/h$) with a tendency to peak when approaching 1. The variance of conductance is clearly scale invariant at $p_c$ (up to weak finite-size corrections) and exhibits a double-peak feature reminiscent to the one observed in the QH transition~\cite{cho1997conductance}. These properties are qualitatively similar to those of critical distributions in disordered systems~\cite{galstyan1997localization,arovas1997real,jovanovic1998conductance,slevin2000topology,cain2001integer,kramer2005random,schweitzer2005universal}. 

When decreasing density towards the threshold I (or VI), the CD acquires a peak near zero conductance (Fig.~\ref{fig:dists}(b)), ultimately becoming a delta peak when density approaches the percolation threshold of the lattice (Fig.~\ref{fig:dists}(c)). At the threshold, the CD can be expressed as $f_{p_c}(\sigma)=(1-\alpha)\delta(\sigma)+\alpha h(\sigma)$ with $0<\alpha<1$ denoting the fraction of connected lattice configurations. Here $h$ is a normalized distribution which controls the finite conductance part. The striking appearance of the low-conductance peak is a consequence of the vicinity of the percolation threshold where 50\% of the configurations become disconnected with vanishing conductance. In the thermodynamic limit, the zero-conductance delta function will vanish above the percolation threshold but unavoidably leaves behind a non-singular low-conductance peak. Interestingly, the double-peak feature of the variance near $p_c$ is not observed at low densities.

The CDs for $\sigma_{xy}$ are shown in Fig.~\ref{fig:dists}(d). At high densities, the distributions of $\sigma_{xy}$ and $\sigma_{xx}$ show strong qualitative differences as in the QH systems~\cite{kramer2005random}. However, when approaching the threshold I (or VI), both distributions acquire a similar form. This further reinforces the fact that the critical behaviour at low and high densities is dominated by distinct mechanisms. Since the distribution functions in discrete and continuum geometries (including the exponential hopping model studied in Sec.~VI in the SI) lead to qualitatively similar conclusions, we identify the low-conductance peak as a generic characteristic of ATC.  

Together, the conductance scaling and the CDs provide a compelling evidence that the remarkable characteristics of ATC arise from the interpolation of a geometric-type transition at low and conventional localization-type transition at high densities. Near the threshold I (or VI), the localization length exponent and the CD functions are consistent with the picture that the critical wave functions essentially reflect the geometry of the underlying lattice. As the density is increased, these signatures evolve smoothly to a different form and the CDs share qualitative features of QH systems.

\emph{Discussion.---}The discovered features of ATC, while striking in the light of the literature accumulated during the last four decades, do not contradict the conclusions of the conventional scaling theory in disordered systems. Despite the superficial similarity, the essential features of the transition on random lattices with varying density are not captured by disordered models on regular geometries. Varying $p$ introduces a variable length scale $l\propto|p-p_c^{cl}|^{-4/3}$ in the system, where $p_c^{cl}$ is the percolation threshold of the lattice. When $p>p_c^{cl}$, this scale characterizes the linear size of randomly placed holes in the lattice. The geometry near $p=1$ is described by dense system with isolated vacancies, while in the limit $p\to p_c^{cl}$ the holes on a lattice diverge $l\to \infty$, leaving only a fractal critical cluster at $p_c^{cl}$. In the dense system the geometric correlations have very short range while they diverge at $p_c^{cl}$. Since the nature of correlations in the disordered systems are known to affect the universality class of the transition~\cite{sandler2004correlated,cain2001integer}, it is natural to consider the variable scale $l$ of the geometric fluctuations as the source of the non-universality. Interestingly, when some aspects of geometric fluctuations were recently implemented in disordered models, the critical exponents were observed to exhibit variation~\cite{klumper2019,gruzberg2017geometrically,chen2019random}. We speculate that the reason for that behaviour reflects the nonuniversal scaling established in the present work. 

The present work has fundamental ramification on the rapidly growing field of amorphous topological systems. The first experimental realizations of elemental and artificial amorphous topological systems have recently become accessible. Thus, it is plausible that the remarkable aspects of ATC can be probed in experiments soon. A comprehensive characterization of ATC can be carried out by probing systems at different densities or variable geometric fluctuations. This could be most naturally carried out in designer systems~\cite{poyhonen2018amorphous,zhou2020amorphous} where density of lattice sites or geometry of the lattice can be easily controlled. The present work also opens many new lines of research. For example, what are the consequences of ATC on other symmetry classes and dimensions such as recently studied amorphous Bi$_2$Se$_3$~\cite{corbae2019evidence}? How do the statistical properties of wave functions reflect the ATC? How are the dynamical properties affected? What new features will quenched disorder add to ATC? These questions will be studied in the future.

\emph{Summary.---}In this work we studied critical transport in Chern insulators with random geometry and discovered remarkable amorphous scaling behaviour. In striking contrast to conventional expectations, the critical exponents and critical conductance distributions characterizing the transition are strongly nonuniversal. Our results indicate that by varying density without affecting symmetries, amorphous topological phase transitions interpolate between a geometric percolation-type and Anderson localization-type transitions. The discovered nonuniversal scaling is a generic feature of amorphous topological matter, indicating striking departure from conventional topological systems.

\bibliography{amorphous}
\bibliographystyle{apsrev4-1}

\pagebreak
\widetext

\begin{center}
\Large SUPPLEMENTAL INFORMATION to ``Criticality in amorphous topological matter --- beyond the universal scaling paradigm''
\end{center}


\section{Evaluating configuration-averaged conductances}
In this section we provide details of our method of calculating configuration-averaged conductances in percolation-type random problems studied in the main text. By applying the analytical connection between the number of occupied sites and the occupation probability (or intensity), our technique results in strongly suppressed statistical fluctuations compared to a conventional approach with the same number of configurations.

\subsection{Discrete random geometry}

A central object in the discrete random lattice case is the mean conductance as a function of the single-site occupation probability $p$. With the conventional approach to percolation problems, one fixes the occupation probability $p$ and calculates the configuration-averaged conductance $\braket{\sigma(p)}$ directly. The novelty employed in our work is to first calculate the conductances $\braket{\sigma(n)}$ as a function of the occupied lattice sites $n=0,\dots,N=L^2$. Since the probability to have $n$ occupied lattice sites is given by the binomial distribution, we obtain the quantity $\braket{\sigma(p)}$ as a convolution with the binomial distribution,
\begin{equation}\label{trafo_binomial}
\braket{\sigma(p)} = 
\sum_n {N\choose n} p^n (1-p)^{N-n} \braket{\sigma(n)} \equiv
\sum_n P(n|p) \braket{\sigma(n)} .
\end{equation}
In particular, this method allows us to choose the desired  probability $p$ \emph{after} the numerical calculations instead of fixing it a priori. At the transition region, the average conductances $\braket{\sigma(n)}$ are typically calculated with a few times $10^3$ different configurations by employing the KWANT package. Since a given $\braket{\sigma(p)}$ is generally calculated using several hundred different $n$ corresponding to the width of the binomial distribution, the number of random configurations contributing to a single point of $\langle\sigma(p)\rangle$ in the main text is up to several times $10^5$. 

A standard way to estimate the uncertainty of a random variable $X$ is to use the standard deviation $\sqrt{\var (X)}$.
Since we are interested in estimating the error bars for the mean conductance, we can compute its variance from the variance of the conductance itself, using $\var \braket{\sigma(n)} = \frac{\var \sigma(n)}{N_\mathrm{samples}}$.
Using Eq.\ \eqref{trafo_binomial} and the properties of variance for a sum, we can deduce the variance of the mean conductance for a given $p$ as
\begin{equation}\label{var_p}
\var \braket{\sigma(p)}
= \var \left( \sum_n P(n|p) \braket{\sigma(n)} \right)
= \sum_n P(n|p)^2 \frac{\var \sigma(n)}{N_\mathrm{samples}} .
\end{equation}
The power of our the method is manifested in the variance formula. The factor $P(n|p)^2\ll 1$ strongly reduces the statistical fluctuations compared to direct $p$ sampling variance which would be comparable to $\frac{\var \sigma(n')}{N_\mathrm{samples}}$ with $n'$ corresponding to the peak value of $P(n|p)$. The relative suppression in the studied systems can even be a few orders of magnitude.   

The values of many different $\braket{\sigma(p)}$ can be inferred from the same $\braket{\sigma(n)}$ data as statistically dependent variables. For the purpose of the scaling analysis in Sec.~II, we need the statistical covariance matrix of the conductances at different $p$ points.
For clarity, denote now the average conductance with an overline $\braket{\sigma} \equiv \widebar{\sigma}$.
For a finite statistical sample $N_\mathrm{samples}$ for all $\widebar{\sigma}(n)$, the covariance between average conductances for different occupation probabilities $p$ is given by
\begin{align}\label{covar_p}
C_{12}=\cov \Big( \widebar{\sigma}(p_1),\widebar{\sigma}(p_2) \Big)
&= \braket{\widebar{\sigma}(p_1)\widebar{\sigma}(p_2)} - \braket{\widebar{\sigma}(p_1)}{\braket{\widebar{\sigma}(p_2)}} \nonumber \\
&= \sum_{n_1,n_2} P(n_1|p_1) P(n_2|p_2)
\Big( \braket{\widebar{\sigma}(n_1)\widebar{\sigma}(n_2)} - \braket{\widebar{\sigma}(n_1)}{\braket{\widebar{\sigma}(n_2)}} \Big) \nonumber\\
&= \sum_n P(n|p_1) P(n|p_2)
\Big( \braket{\widebar{\sigma}(n)^2} - 
 \braket{\widebar{\sigma}(n)}^2 \Big) \nonumber\\
&= \sum_n P(n|p_1) P(n|p_2) \var \widebar{\sigma}(n) \nonumber\\
&= \sum_n P(n|p_1) P(n|p_2) \frac{\var \sigma(n)}{N_\mathrm{samples}} ,
\end{align}
where we have used the multiplicativity of the mean for independent random variables when $n_1 \neq n_2$. As it should, the covariance of $\widebar{\sigma}(p)$ with itself returns the variance given in Eq.\ \eqref{var_p}.

\subsection{Continuum random geometry}

In the case of continuum systems, we consider continuum percolation-type geometries and focus on constant hopping
in a finite radius $r \le R$. In Appendix \ref{app:cont_exp}, we also present the conductance distributions for hopping which decays exponentially $\sim e^{-r/\eta}$.
In contrast to the lattice case considered above, there is no underlying structure where lattice sites would be placed. Instead the analysis proceeds as follows:
Heuristically, if an area of size $L^2$ contains $n$ particles with a hopping radius of $R$, we can give the density as the ratio of the sum of disks to the entire system,
$\rho = \frac{n \pi R^2}{L^2} = \lambda \pi R^2$, where $\lambda = n/L^2$ is the particle number per system size. Turning this around, if we fix $\lambda$ in a large system, a subsystem of size $L^2$ contains, on average, $n \approx \lambda L^2$ particles (with fluctuations about the mean).
The lattice points are located either within the confines of the subsystem, or not, independently of all other points. This describes a Poisson distribution with an intensity parameter $\lambda' = \lambda L^2$.
The probability to find $k$ lattice points in our system is then
\begin{equation}
P_{\lambda}(k) = \frac{\lambda'^k e^{-\lambda'}}{k!}, \quad k = 0,1,2,...
\end{equation}
The variable $\lambda$ is now allowed to take on any real positive value. Thus we have in the continuous variable $\rho = \lambda \pi R^2$ the continuum version analogue of the percolation probability $p$ of the lattice case; it describes a distribution for the number of lattice sites $n$, and is peaked at $\rho = \frac{n\pi R^2}{L^2}$.
Unlike here, continuum percolation problems are conventionally studied with disks where contact does not require the center of another one to be in the circle of influence, but only requires overlap of the circles themselves. Hence the conventional variable is the \emph{filling fraction} $\eta$, related through $\rho = 4\eta$.

For a given system size $L$ and hopping radius $R$, the goal is now to calculate the configuration-averaged conductances $\braket{\sigma(n)}$ for a fixed number of lattice points $n$.
In analogy to the lattice case above, the mean conductance as a function of the percolation intensity $\rho$ is then obtained as a convolution with the Poisson distribution,
\begin{equation}\label{trafo_Poisson}
\braket{\sigma(\rho)} = 
\sum_n P_\lambda(n) \braket{\sigma(n)} .
\end{equation}
The analysis of the uncertainty of the mean conductance then closely follows the lattice case; Eqs.\ \eqref{var_p}-\eqref{covar_p} for the (co)variance look identical, except with the appropriate probability distribution $P_\lambda(n)$ given here.
For exponential hopping $\sim e^{-r/\eta}$, the Poisson distribution stays the same, but the Poisson intensity is now expressed in the units of exponential decay length $\eta$ which is the natural length scale of the problem.

The conductances for the continuum geometry are calculated using the Green's function method previously used to study thermal conductance of amorphous topological superconductors \cite{poyhonen2018amorphous}.
The Green's functions
$G_{r,a}^{-1}(E) = E - H - \Sigma_{r,a}$ 
have a self-energy term originating from the coupling to the electronic leads, and is given by
\begin{equation}
\Sigma(E,m,n) =
\frac{1}{L_y/a+1} \left(\frac{t_C}{t_L}\right)^2
\sum_k \sin(kma) (\varepsilon - i\sqrt{4t_L^2-\varepsilon^2}) \sin(kna),
\quad |\varepsilon|<2t_L,
\end{equation}
where $ma,na$ are the $y$-coordinates of sites connected to the leads, and $t_C$ and $t_L$ are the hopping parameter between the lead and the Chern glass, and the hopping parameter within the lead, respectively. The leads are modelled as square lattice structures, and we have defined $\varepsilon = E - 2t_L\cos(ka)$.
In the above expressions, $k$ takes values $k(j) = \frac{j\pi}{a(N_y+1)}$ for $j = 1\ldots N_y$, where $N_y$ denotes the number of fixed sites on each side of the scattering region. The  conductance through the Chern glass (in the units of $e^2/h$) is then given by
\begin{equation}
g = \mathrm{Tr}\left[\Gamma_L G_r^{LR} \Gamma_R G_a^{RL} \right].
\end{equation}
Here, we have defined $\Gamma_{L,R} = -2\IM \Sigma_{L,R}$ as well as the matrices $G_{r,a}^{LR}$ and $G_{r,a}^{RL}$, the latter of which are the subblocks of the Green's function connecting the leftmost and rightmost edges of the sample.


\section{Scaling analysis and error bounds} 

 In this section we discuss the scaling theory developed for topological localization transitions in integer quantum Hall (IQH) systems. Close to criticality, a thermodynamic quantity of interest, e.g. the configuration-averaged conductance, follows the scaling relation $\braket{\sigma}$ = $F[L^{1/\nu} \zeta_1(p),L^{y}\zeta_2(p)]$, where $F$ defines a two-parameter scaling function. The scaling relation can be expanded in its arguments~\cite{huckestein1995scaling,slevin1999corrections,amado2011numerical,obuse2012finite} as
\begin{equation}\label{eq:tp-scaling}
\braket{\sigma} =\sum^{n}_{k=0} \zeta_2^{k}\,L^{ky}\,F_k[L^{1/\nu} \zeta_1(p)],
\end{equation}
where $\nu$ is the critical localization length exponent, $p$ (the filling probability of lattice sites) is the parameter controlling the localization behavior and the irrelevant exponent $y<0$ characterizes power law corrections. Generally, the type of finite-size corrections and the predicted asymptotic values of thermodynamic quantities are affected by the effective range of system sizes studied in a numerical simulation. In principle, one could also consider logarithmic-like corrections \cite{amado2011numerical,puschmann2019integer,zhu2019localization,nuding2015localization}, which stands beyond the scope of common field theories developed for IQH phase transition. Each $F_k$ in Eq.~\ref{eq:tp-scaling} can be further expanded up to the order $n_R$:
\begin{equation}
F_k[L^{1/\nu} \zeta_1(p)] = \sum^{n_R}_{m=0} \zeta_1^m\, L^{m\nu} F_{km}.
\end{equation}

Moreover, to account for the possible non-linearities in the scaling variables $\zeta_1(p), \zeta_2(p)$, we expand them in the dimensionless probabilities $p-p_c$:
\begin{equation}
 \zeta_1(p) = (p-p_c)+\sum^{m_R}_{i=2} b_i\,(p-p_c)^i \quad\text{and}\quad \zeta_2(p) = 1+\sum^{m_I}_{j=1} c_j\,(p-p_c)^j.
\end{equation}

In the theory of IQH transition, $\zeta_1(p), \zeta_2(p)$ are associated with the deviations from the fixed-point values of $\sigma_{xy}$ and $\sigma_{xx}$, respectively. Here we note that the flow towards the fixed point (along the line $\zeta_1=0$) is described by the single-parameter scaling relation~\cite{schweitzer2005universal,kramer2005random,wang1996critical,wang1998scaling}:
\begin{equation}\label{eq:xx-scaling}
\braket{\sigma_{xx}(L)}_c = F[0,L^{y}\zeta_2] =\,\braket{\sigma_{xx}}_c + F'(0,0)\, \zeta_2\, L^{y},
\end{equation}
such that $y$ can be found by studying the finite-size dependence of the critical longitudinal conductance. We employed the scaling approach \eqref{eq:tp-scaling} for systems of the size $L\times L$, which can be used conveniently to fit the data from both the strip and cylinder geometries.

For the fitting purposes we consider a nonlinear least-square minimization, using a trust region reflective method.
The estimates of the model parameters are obtained by minimizing the general cost function given by:
\begin{equation}\label{eq:chi_sq}
\chi^2= \sum_{ij}R_i\,C^{-1}_{ij}\,R_j,
\end{equation}
where $R_i=\sigma(L,p_i) - F(L,p_i)$ is the residual function and $C_{ij} = \cov(\sigma(p_i),\sigma(p_j))$ is the (absolute) statistical covariance associated with the mean conductance data points, and the sum extends over different system sizes $L$~\cite{michael1994fitting}. To asses the quality of the fitting we considered several standard criteria developed in this regard, specifically the generalized $\chi^2$-test and confidence intervals on the estimated parameters~\cite{press2007numerical}. For an ideal fit, $\chi^2_{r}=\chi^2/N_{dof}$ is expected to be 1, where $N_{dof}$ is the difference between the number of data points and fit parameters $N_{dof}=N_d-N_p$. By calculating the covariance matrix on the sets of obtained model parameters, standard deviations are estimated. Independently, we further considered the $68\%-99\%$ confidence intervals, observing well-behaved and consistent results. We have ensured for the best fit to be stable against higher orders of the expansion; after that the critical parameters obtained from the simplest model in a specified interval are reported.

\begin{figure}[htp]
    \centering
    \includegraphics[width=.79\linewidth]{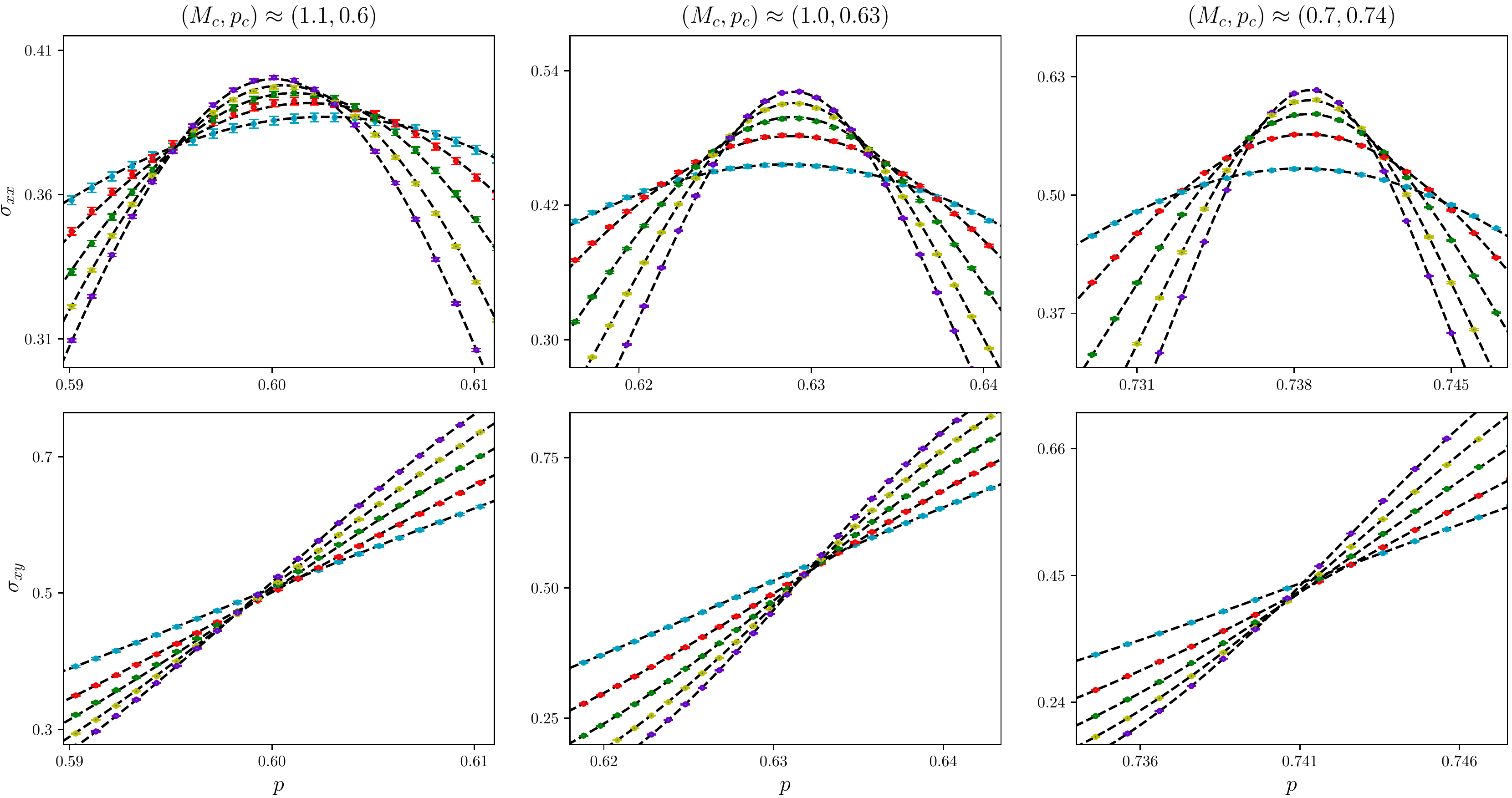}
    \caption{$\sigma(p,L)$ as a function of $p$ for $L=40,60,80,100,120$. The dashed lines result from the best fit.}
    \label{fig:scaling_fit}
\end{figure}

\Cref{tab:pbc_scaling_sq,tab:obc_scaling_sq,tab:obc_scaling_cont} summarize the result of fitting for discrete and continuum geometries and Fig.~\ref{fig:scaling_fit} (and Fig.~2 in the main text) shows examples of the scaling fits. It has been previously observed that the different boundary conditions could affect the estimated critical values, especially the value of the irrelevant exponent~\cite{puschmann2020edge,obuse2010conformal,slevin2000topology}. However, in our work we observe generally consistent values even for the exponent $y$. The results from both open and periodic boundary conditions show excellent agreement for the critical density $p_c$ and the critical exponent $\nu$. 


\begin{table}[t]
\caption{Critical parameters and their uncertainties obtained at different mass parameters $M$ by varying the site-densities $p$ for the longitudinal conductance (cylindrical geometry). $N_d$ and $N_p$ specify the number of data points and fitting parameters used in the fitting procedure, respectively. $\chi^2_{r}$ denotes the value of the reduced chi-squared.}

\begin{tabular}{|P{2cm}||P{2cm}|P{2cm}|P{2cm}|P{2cm}|P{2cm}|P{2cm}|P{2cm}|}
$M$ & $p_c$ & $\sigma^{c}_{xx}$ & $\nu$ & $|y|$& $N_d$ & $N_p$& $\chi^2_{r}$ \\
\hline
1.1 & 0.596(2) & 0.39(2) & 1.35(2) & 0.88(4) & 759 & 14 & 1.02\\
1.0 & 0.629(2) & 0.56(2) & 1.20(2) & 0.67(3) & 676 & 13 & 1.02\\
0.9  & 0.665(2)  & 0.64(3) & 1.11(2) & 0.62(3) & 653 & 12 & 1.02\\
0.7 & 0.739(2) & 0.72(3) & 1.05(1) & 0.54(4) & 640 & 12 & 1.01\\
0.5  & 0.813(2)  & 0.78(3) & 1.01(1) & 0.48(5) & 634 & 12 & 1.01\\
\end{tabular}
\label{tab:pbc_scaling_sq}
\end{table}

\begin{table}[t]
\caption{Same as Table~\ref{tab:pbc_scaling_sq} for the Hall conductance (strip geometry).}
\begin{tabular}{|P{2cm}||P{2cm}|P{2cm}|P{2cm}|P{2cm}|P{2cm}|P{2cm}|P{2cm}|}
$M$ & $p_c$ & $\sigma^{c}_{xy}$ & $\nu$ & $|y|$ & $N_d$ & $N_p$& $\chi^2_{r}$\\
\hline
1.1 & 0.597(2) & 0.49(2) & 1.34(2) & 0.99(4) & 742 & 13 & 1.02\\
1.0 & 0.631(2) & 0.44(2) & 1.19(2) & 0.68(3) & 669  & 13 & 1.01\\
0.9 & 0.667(2)& 0.36(2)  & 1.10(2) & 0.64(4) & 621 & 12 & 1.02\\
0.7 & 0.739(2) & 0.31(3) & 1.05(1) & 0.58(4) & 623 &12 & 1.01\\
0.5  & 0.813(2)  & 0.29(3) & 1.01(1) & 0.51(5) & 638 & 12 & 1.01\\
\end{tabular}
\label{tab:obc_scaling_sq}
\end{table}

\begin{table}[h]
\caption{Critical parameters and their uncertainties for the Hall conductance in the continuum model.}
\begin{tabular}{|P{2cm}||P{2cm}|P{2cm}|P{2cm}|P{2cm}|P{2cm}|P{2cm}|P{2cm}|}
$M$ & $\rho_{c}$ & $\sigma^{c}_{xy}$ & $\nu$ & $|y|$ & $N_d$ & $N_p$& $\chi^2_{r}$\\
\hline
0.85 & 4.73(2) & 0.31(2) & 1.32(3) & 1.68(5) & 512 & 13 & 1.02\\
-0.5 & 6.67(2) & 0.46(2) & 1.10(3) & 1.07(9) & 668 & 12 & 1.02\\
-1.0 & 7.67(2) & 0.44(3) & 1.17(3) & 0.93(8) & 516 & 12 & 1.02\\
-1.5 & 8.67(2) & 0.45(2) & 1.07(2) & 1.51(8) & 600 & 13 & 1.02\\
\end{tabular}
\label{tab:obc_scaling_cont}
\end{table}

\section{Two-terminal geometry with open transverse boundary conditions and Hall conductance}

\begin{figure}[H]
    \centering
    \includegraphics[width=.45\linewidth]{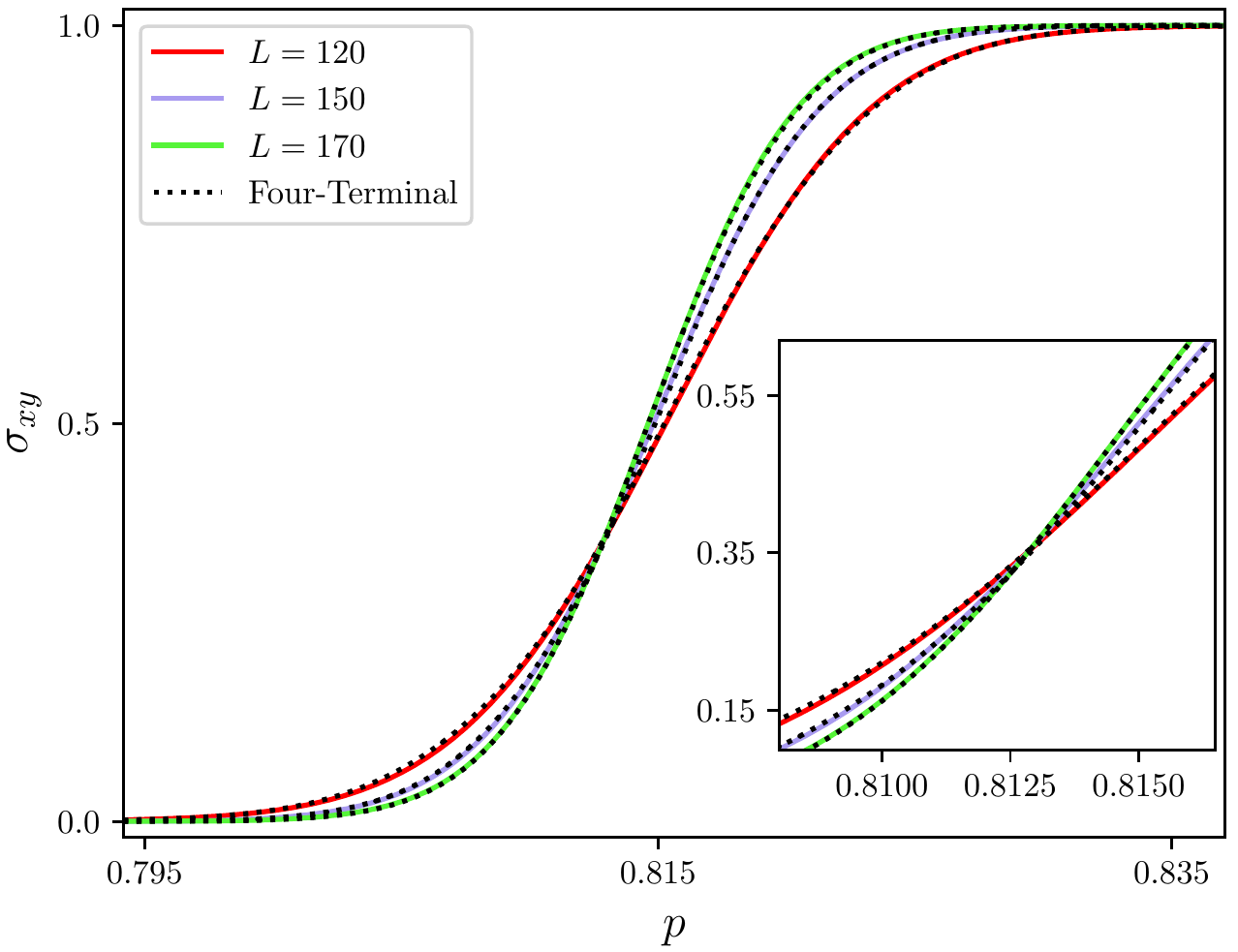}
    \caption{Comparison of the conductances obtained from two-terminal setup with open transverse boundary conditions (solid lines) and four-terminal setup (dotted lines) at $M=0.5$. Inset displays a zoom-in plot in the vicinity of the transition point $(p_c,M_c)\approx(0.813,0.5)$.}
    \label{fig:two_four_term}
\end{figure}

In practice, conductance calculations are performed in square samples in a two-terminal geometry. Employing four or six-probe geometries are inconvenient for carrying out random configuration averages. While the two-terminal conductances corresponding to periodic and open boundary conditions in the transverse region are perfectly good scaling variables, it is interesting that these actually correspond to longitudinal and Hall conductivity. While the correspondence between the cylindrical geometry and longitudinal conductivity is intuitive, the one between the Hall conductance and open boundary conditions is less clear. The purpose of this section is to numerically illustrate the correspondence between the conductances calculated  from two-terminal setup with strip geometry and four-terminal setup which defines the experimentally accessible Hall conductance $\sigma_{xy}$~\cite{datta1997electronic,shun2018topological}. As displayed in Fig.~1 in the main text, in the case of four-terminal measurement, a current $I$ is applied through the two large terminals and the voltage difference $\Delta V$ between the floating side probes is measured. Thus, for an $L\times L$ system we have $\sigma_{xy}=|I/\Delta V|$. As shown in Fig.~\ref{fig:two_four_term}, we observe consistent behavior for the two- and four-probe measurements. Here the data is averaged over 100-500 different realizations for each number of occupied lattice sites $n$.


\section{Conductance distribution functions}

A major part of our analysis is the calculation of the critical conductance distribution functions along the phase boundary.
As in the conductance calculations described in Sec.~I, we carry out configuration averages for different number of lattice sites and employ the analytic connection to express quantities as a function of density. For a fixed number of occupied lattice sites $n$, the distribution function $f_n(\sigma)$ has the following properties:
(1) It is normalized to unity, $\int f_n(\sigma) d\sigma = 1$,
(2) the mean of the conductance is given by the first moment of the distribution,
$\braket{\sigma}_n   
= \int f_n(\sigma) \sigma d\sigma$, and
(3) the variance of the conductance is given by the second central moment, $(\delta\sigma_n)^2 = \int f_n(\sigma) (\sigma-\braket{\sigma}_n)^2 d\sigma.$ 
For a fixed occupation probability $p$, there are $n$ occupied lattice sites with probability $P(n|p)$, given by the binomial distribution.
Thus, the distribution function of the conductance in terms of $p$ is given as a weighted sum over the distribution functions for the number of lattice sites $n$,
\begin{equation}
f_p(\sigma) = \sum_n P(n|p) f_n(\sigma) ,
\end{equation}
which is also normalized to unity.
The first moment of the distribution and the mean conductance is then given by a convolution with the binomial distribution as
\begin{equation}
\label{eq:cond_mom1_p}
\braket{\sigma}_p =   
\int f_p(\sigma) \sigma d\sigma = \sum_n P(n|p) \int f_n(\sigma) \sigma d\sigma = \sum_n P(n|p) \braket{\sigma}_n. 
\end{equation}

It should also be noted that in the continuum case, there is no underlying lattice, and the mean of the conductance is analogously given by an equation similar to Eq.\ \eqref{eq:cond_mom1_p}, except with the probability $P(n|\rho)$ given by the Poisson distribution.
The variance of the conductance is then given by 
\begin{equation} \label{eq:cond_mom2_p}
(\delta\sigma_p)^2 = 
\int f_p(\sigma) (\sigma - \braket{\sigma}_p)^2 d\sigma = 
\int f_p(\sigma) \sigma^2 d\sigma - \braket{\sigma}_p^2 = 
\sum_n P(n|p) \braket{\sigma^2}_n - \left( \sum_n P(n|p) \braket{\sigma}_n \right)^2 .
\end{equation}
%
The critical conductance distribution $f_{p_c}(\sigma)$ can be calculated straightforwardly when the critical density $p_c$ is known. To further characterize the conductance fluctuations for random lattices, the second moment of the conductance $(\delta\sigma_{p})^2$ is studied where $\delta\sigma_p$ is the root-mean-square for a given large conductance ensemble. The second moment shares common scaling behavior with $\sigma (L,p)$. \cite{galstyan1997localization,arovas1997real,jovanovic1998conductance,cain2001integer,romer2002percolation,kramer2005random,schweitzer2005universal,cho1997conductance}


\section{Mass vs density driven phase transitions}
\begin{figure}[h]
\centering
\includegraphics[width=0.45\textwidth]{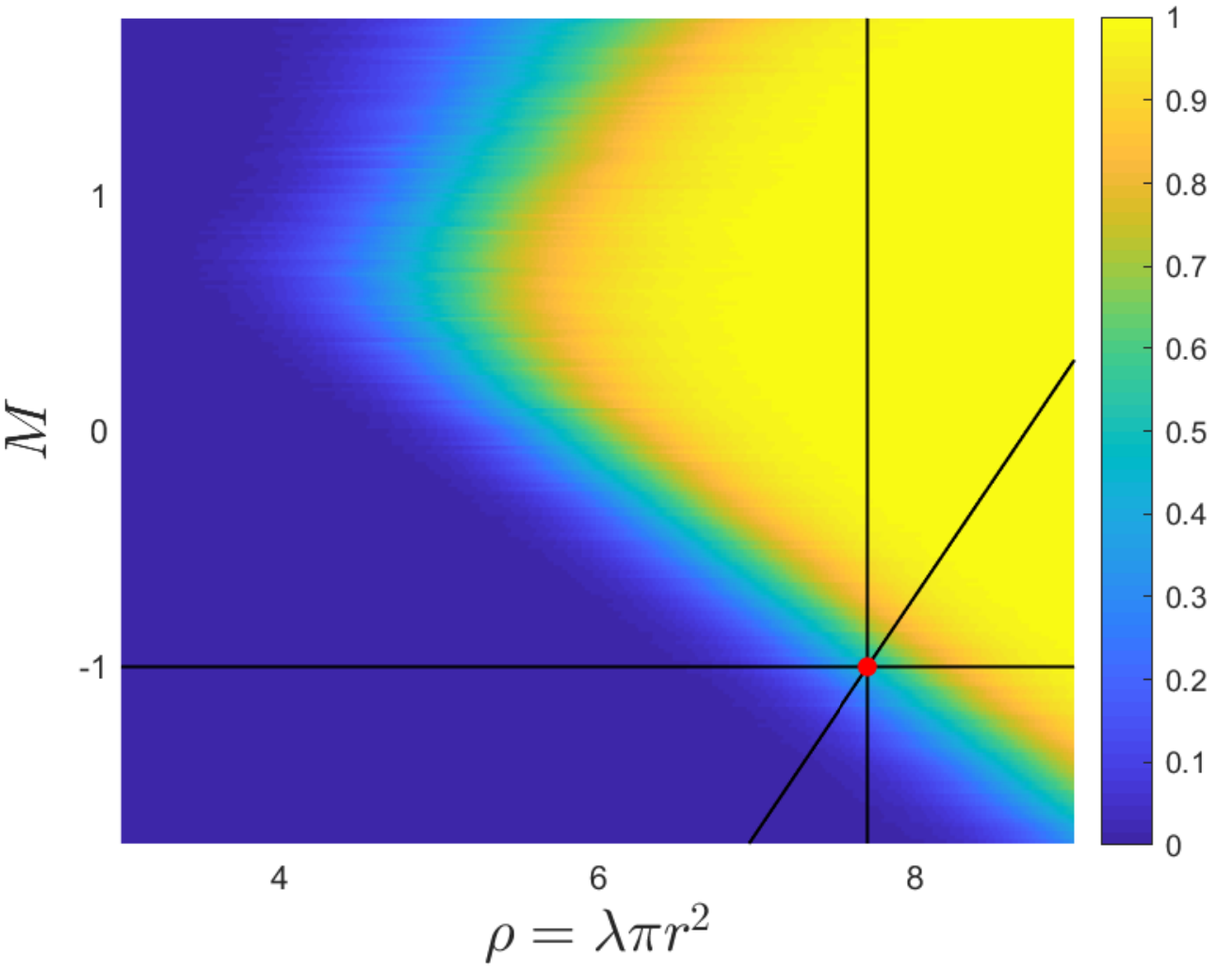}
\includegraphics[width=0.42\textwidth]{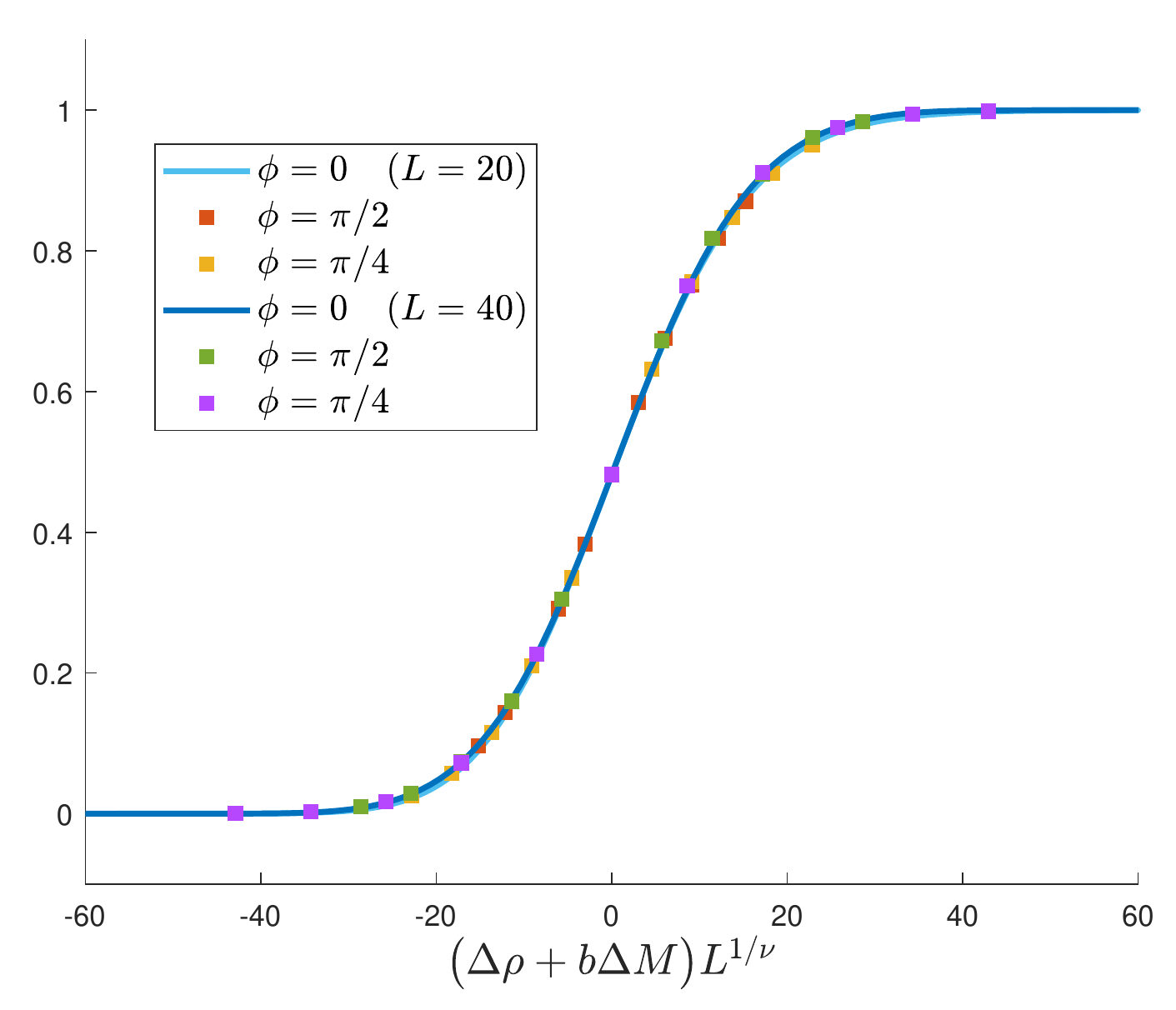}
\caption{
The phase boundary can be crossed through a transition point in different directions. In the left figure, the black lines through the red critical point $(\rho_c,M_c) \approx (7.7,-1)$ show some example directions compared here, with $\phi=0$ corresponding to the horizontal direction of fixed $M$. In the right figure, the conductance curves along these three directions are given for two system sizes, collapsed with the critical exponent $\nu=1.17$ found in Table \ref{tab:obc_scaling_cont}. The numerical constant appearing in the localization length expression is $b\approx 2.0$ for this transition point.
}
\label{fig:cont2D_withAngles}
\end{figure}

In the main text we consider transitions between the trivial and nontrivial phases with fixed mass parameter $M$ by varying the occupation probability $p$ or the continuum analogue $\rho$. The transition can be driven by different mechanisms as well, for example, by varying the mass at constant density or varying both in some fashion through the phase boundary point $(P_c,M_c)$ as depicted in Fig.~\ref{fig:cont2D_withAngles}.
The natural expectation is that the localization length exponents $\nu$ in the different directions would agree. Thus, the localization divergence for $p$ and $M$ driven transitions $\xi\propto|p-p_c|^{-\nu}$, $\xi_M\propto|M-M_c|^{-\nu}$ would have the same $\nu$. Due to the numerous unexpected departures from the orthodox picture, we want to check this explicitly.   

As explained in Sec.~I, our method of evaluating configuration averages is tailor-made for the density-driven transition. Thus, the results for other transitions presented here have a more qualitative nature. 

In Fig.~\ref{fig:cont2D_withAngles} (left) we show three transitions characterized by different crossings through the boundary at the critical point $(\rho_c,M_c) \approx (7.7,-1)$. The different transitions can be labelled by the angle $\phi$ between the $\rho$ axis and the transition direction. In Fig.~\ref{fig:cont2D_withAngles} (right) we see that the conductance data obtained from the mass-driven transition with fixed $\rho$ corresponding to $\phi=\pi/2$ collapses excellently on top of the density-driven transition ($\phi=0$) with the common critical exponent $\nu$. This also implies that the transition can be characterized by a diverging localization length defined by $\xi\propto| \rho-\rho_c +b(M-M_c)|^{-\nu}$, where $b$ is a numerical constant. This is further confirmed by considering the transition where both $\rho$ and $M$ are simultaneously varied so that $\phi=\pi/4$. Again, as indicated in Fig.~\ref{fig:cont2D_withAngles} (right), the conductance data from the resulting transition collapses excellently on the curves corresponding pure $\rho$ or $M$ driven transitions with the common critical exponent $\nu$.

\section{Conductance distribution functions for continuum exponential hopping model}
\label{app:cont_exp}
In the main text we consider discrete and continuum random models where the hopping amplitude is constant inside a fixed radius and strictly zero beyond that. The qualitative comparison of these models clearly illustrate the salient features of amorphous Chern insulators and the discovered amorphous topological criticality. The reason for this section is to explicitly show that the generic features of the amorphous topological transition are also observable for the continuum geometry in the exponential hopping model. The model is defined as Eq.~(1) in the main text with hopping amplitudes $T_{ij}=-\frac12 e^{-r_{ij}/\eta}$ with a continuum percolation geometry.  This confirms that the qualitative findings in the main text are not system-specific, but generic for topological models on random lattices. 

\begin{figure}[h]
\centering
\includegraphics[width=0.32\textwidth]{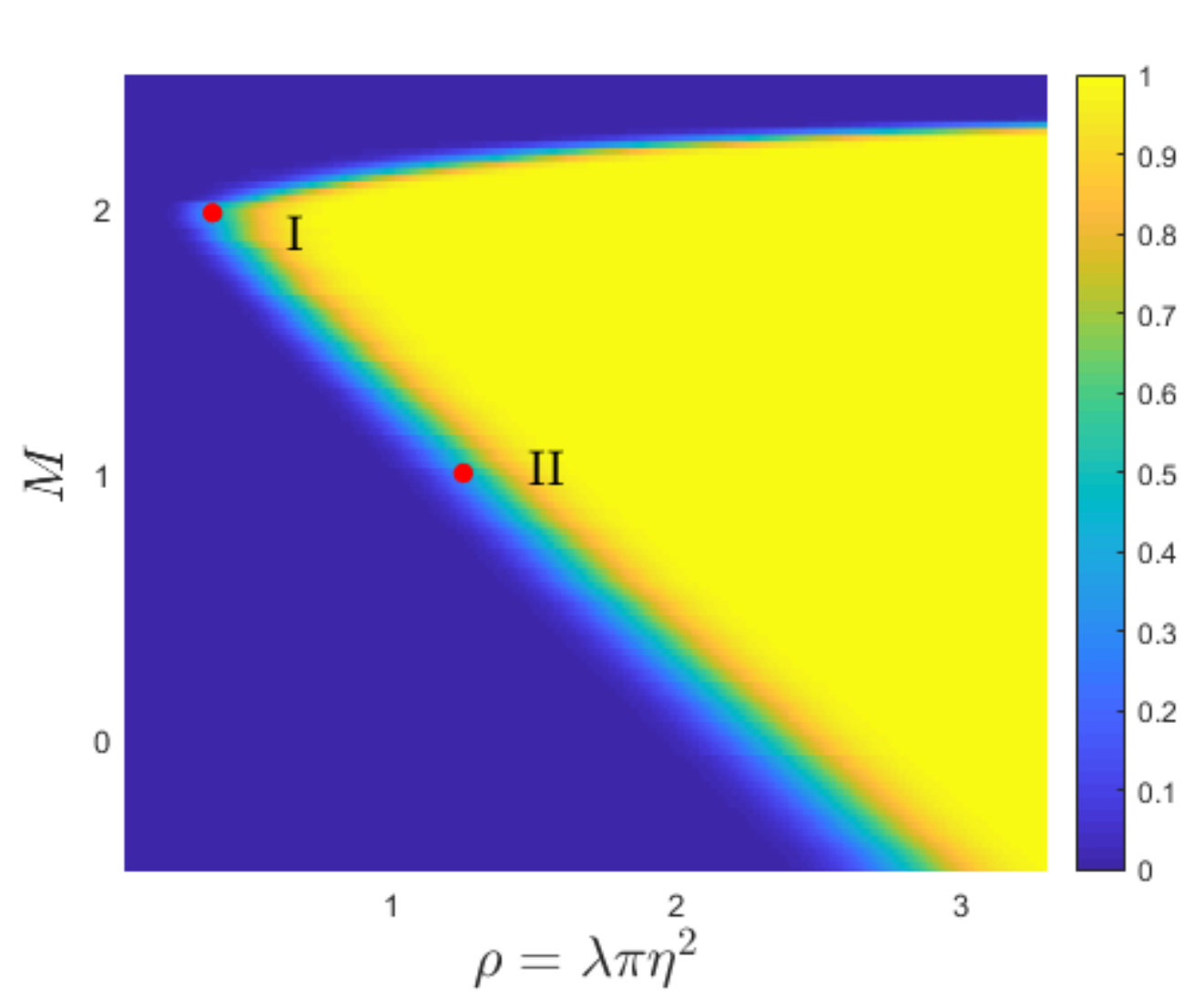}
\includegraphics[width=0.32\textwidth]{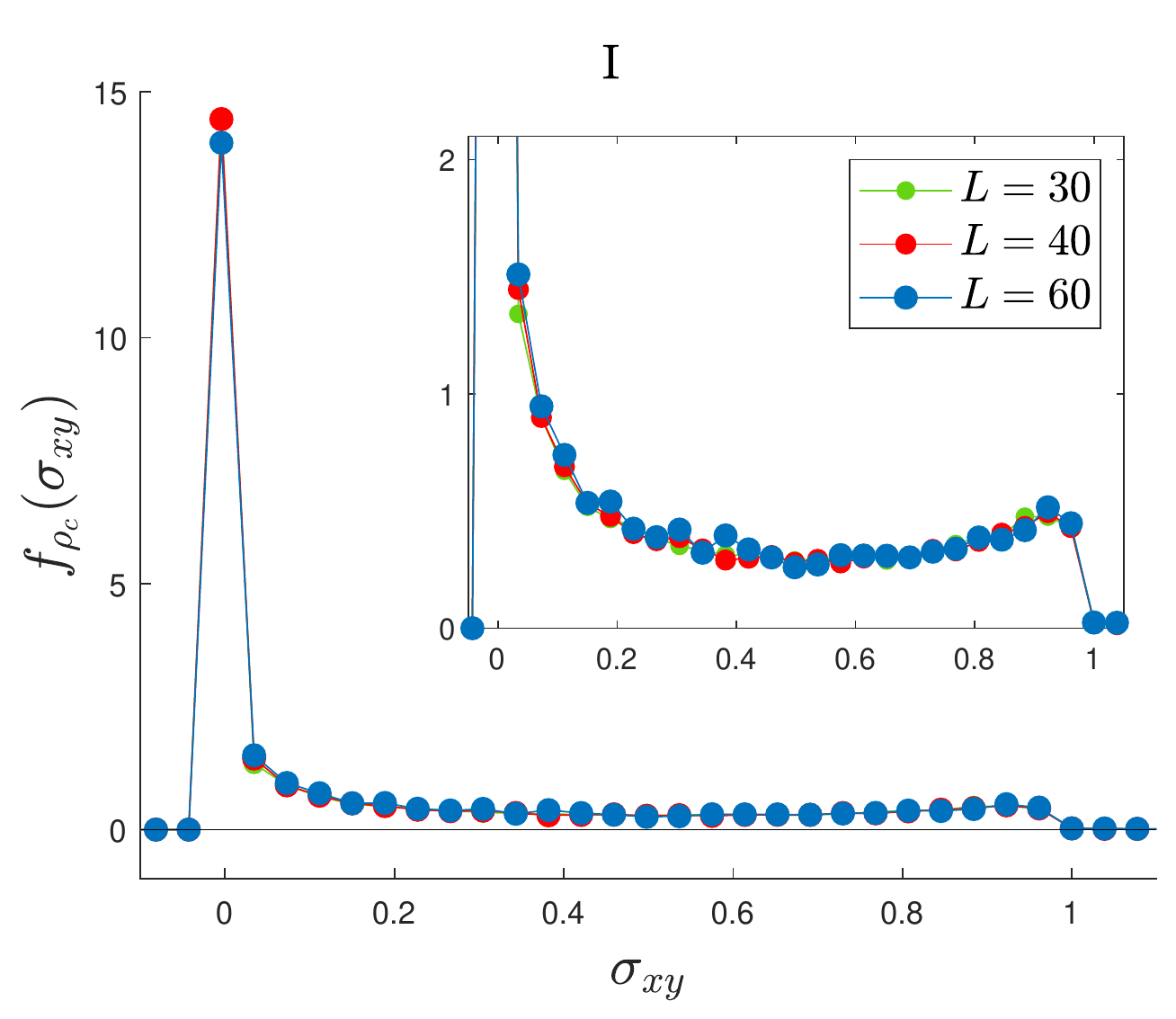}
\includegraphics[width=0.32\textwidth]{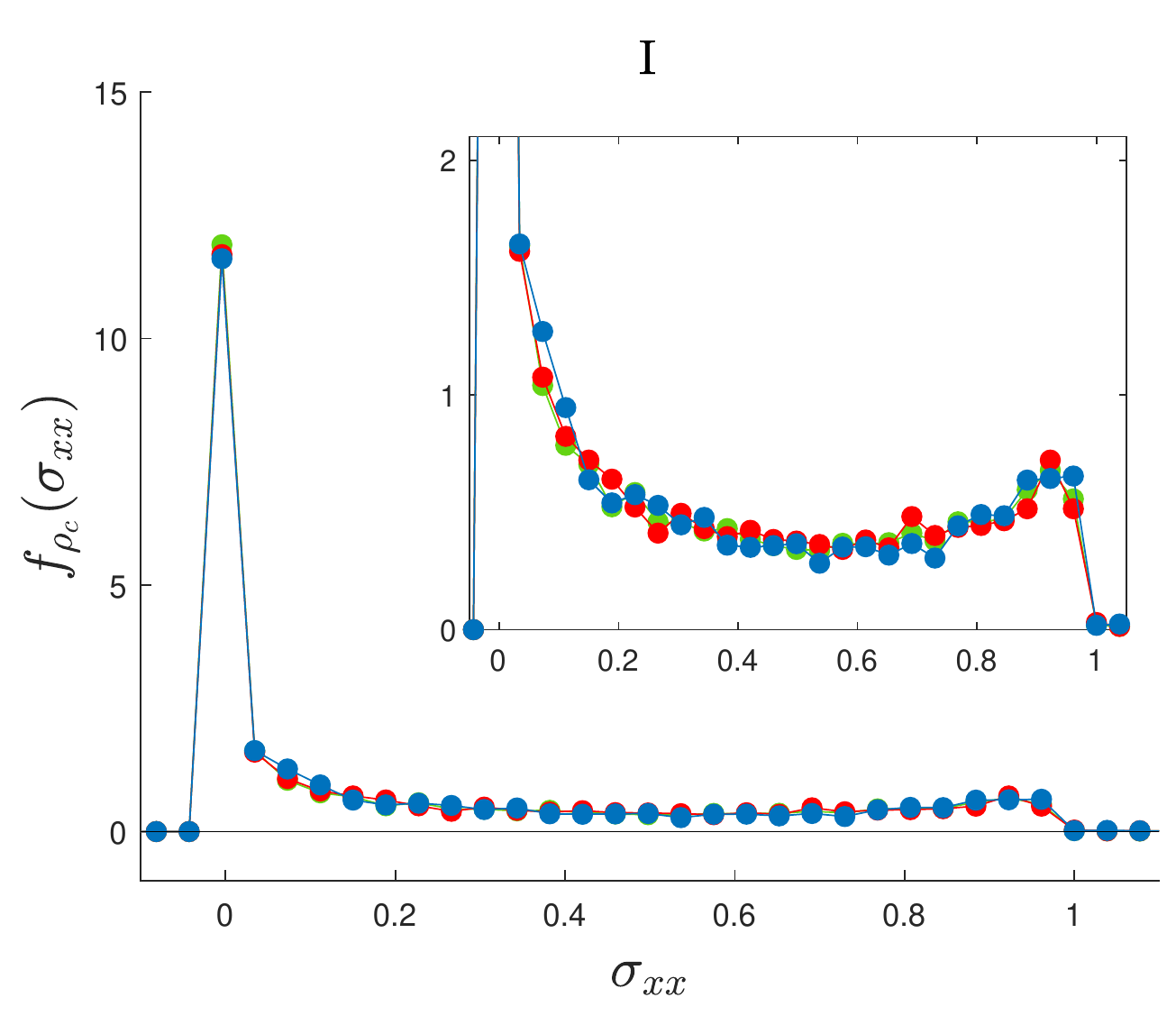}
\\
\centering
\includegraphics[width=0.32\textwidth]{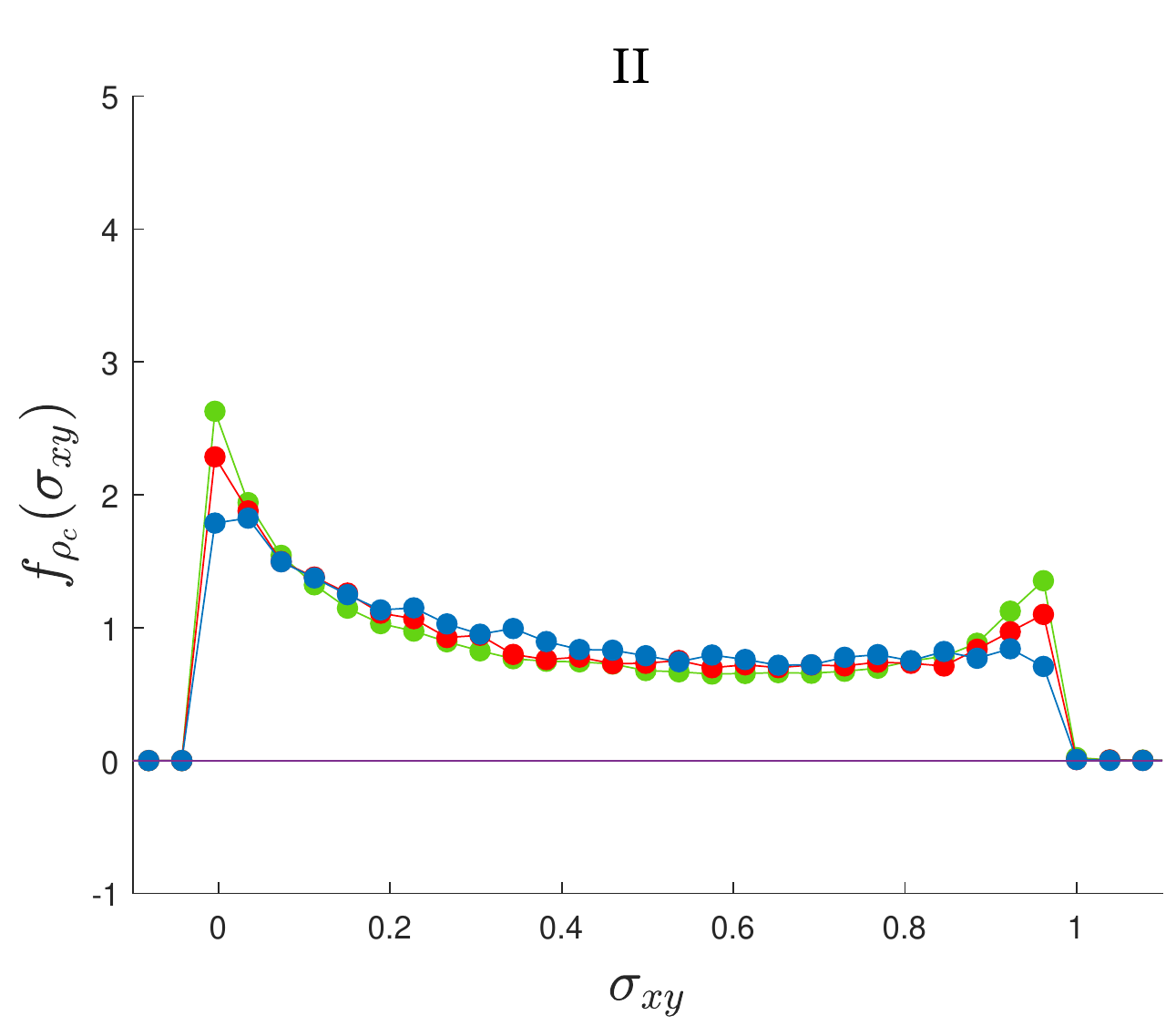}
\includegraphics[width=0.32\textwidth]{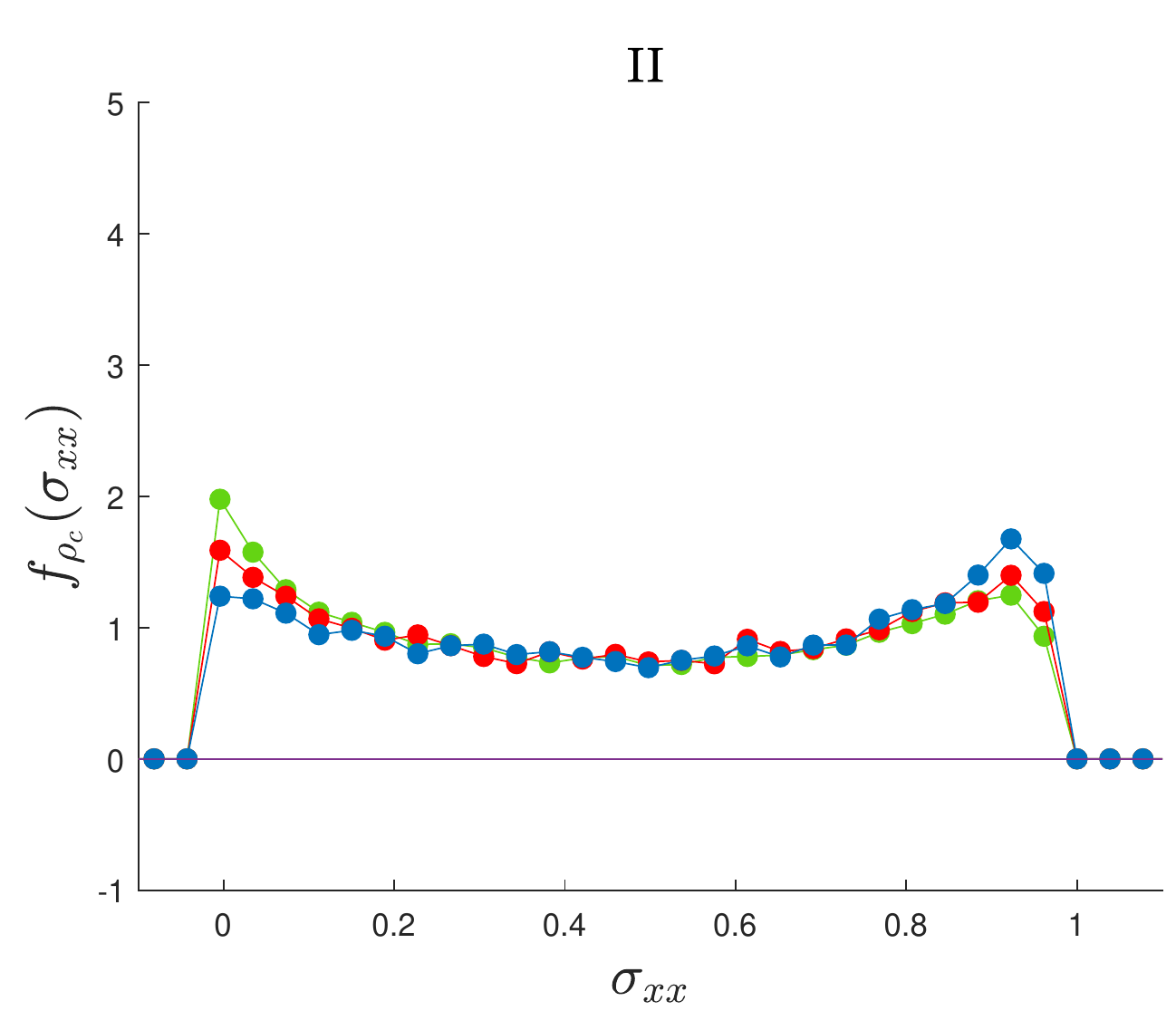}
\caption{Top left: Topological phase diagram for the exponential hopping model on continuum random geometry. Top Center/right: Critical conductance distribution functions at location I. The inset shows a blow-up for the finite conductance part. Bottom left/right: Conductance distribution functions at point II. }
\label{fig:cont_exp}
\end{figure}

The phase diagram of the exponential model is shown in Fig.\ \ref{fig:cont_exp}, where we have also indicated two separately studied points on the phase boundary. In agreement with the models studied in the main text, in the low-density limit (point~I) both $\sigma_{xx}$ and $\sigma_{xy}$ distributions have a similar form with striking low-conductance peaks. Moving towards higher density (point~II), the low-conductance peaks disappear and the conductance distributions acquire a different shape. The critical exponents reflect a similar nonuniversality as observed in the critical conductance distributions.

\end{document}